\documentclass[twocolumn,apj,floatfix]{aastex701}
\pdfoutput=1
\usepackage{amsmath,amssymb,gensymb,color,graphicx}
\usepackage{array}
\usepackage{natbib,hyperref}
\usepackage{multirow}
\usepackage{placeins}
\usepackage{textcomp}

\usepackage{booktabs} 
\usepackage{microtype}
\usepackage{caption}
\usepackage{subcaption}
\usepackage{expl3}

\newcommand\F[1]{\mathcal{F} \left [ #1 \right ]}

\begin{document}
\shorttitle{Atmospheric Noise at Small Angular Scales}

\title{The Effect of Large Apertures on Atmospheric Noise Spectra for Millimeter-wave Telescopes}

\author[0000-0002-5564-997X]{Thomas~W.~Morris}
\email{thomas.w.morris@yale.edu}
\affiliation{Department of Physics, Yale University, New Haven, CT 06511, USA}

\author[0000-0002-9674-4527]{Neelima~Sehgal}
\email{thomas.w.morris@yale.edu}
\affiliation{Department of Physics and Astronomy, Stony Brook University, Stony Brook, NY 11794, USA}

\correspondingauthor{Thomas~W.~Morris}
\email{thomas.w.morris@yale.edu}

\begin{abstract}

Ground-based millimeter and submillimeter telescopes produce astronomical maps that contain a combination of noise inherent to themselves as well as contamination from foregrounds, with the strongest foreground being fluctuating atmospheric emission. While the emission of the turbulent atmosphere itself follows a scale-free power law, the atmosphere as seen by the telescope is modified by the properties of the instrument optics. A naive power-law extrapolation of the atmospheric noise spectrum ignores the blurring effect of the near-field beam, which becomes more pronounced for larger telescope apertures and at higher multipoles. In this work, we use a toy model to derive a semianalytic estimate of the atmospheric noise spectrum that takes into account the effects of near-field optics. We show that it has an appreciable effect for large-aperture telescopes, suggesting that ultra-sensitive, high-resolution experiments are not necessarily limited by the atmosphere on small angular scales. We also present an estimate of the atmospheric knee as a function of the primary aperture, the atmospheric scale height, and the outer scale of turbulence.

\end{abstract}

\section{Introduction}

As more advanced astronomical observatories are deployed to map the sky at higher resolutions and sensitivities, understanding the noise spectrum becomes more crucial. 
While instrument noise can be reduced by deploying more sensitive arrays with additional detectors, atmospheric contamination is heavily correlated across all detectors in a telescope, meaning that two instruments will experience the same atmospheric noise spectrum averaged down at largely the same rate, regardless of their instrument noise properties.
This can lead to diminished returns, as an arbitrarily sensitive instrument may still be limited by atmospheric noise.

Millimeter-wave telescopes are typically diffraction-limited, meaning that they are limited by the Rayleigh criterion in the angular scales that they can resolve. Paradoxically, though, while a larger-aperture telescope has a smaller beam on the sky, it has a larger beam in the near field, as \'etendue is conserved for a diffraction-limited optical system. Even without deriving noise spectra, we can intuit that such a telescope sees less atmosphere at small scales because it averages such fluctuations over a large volume of turbulence and cannot resolve atmospheric anisotropies that are too close to the telescope.\footnote{This is the same reason that interferometers with longer baselines more thoroughly suppress the atmosphere, and the same reason your nose appears blurry in your own field of view.} The implication of this effect is that noise levels forecasted by simulation or extrapolation may be inaccurate on small angular scales if models ignore the near-field optics of the beam.

Near-field beams have been included in earlier atmospheric models and simulations~\citep{church1995predicting,Sayers2010,Errard2015,morris/etal:2022,vanmarrewijk2024maria}, but they can be difficult to study because sub-degree scales are usually dominated by white noise in the time domain. In this paper, we use a toy model to derive a semianalytic estimate of the atmospheric map noise spectrum. We show that this spectrum for instruments designed with larger primary sizes and larger instantaneous fields of view is lower than predicted by the power-law model at small angular scales. While this work focuses on deep, wide-area surveys 
(and is thus most relevant to observations of the CMB), the effect of the near-field beam on atmospheric noise is relevant to any single-dish millimeter-wave telescope, including those making targeted maps of smaller fields.

In Section \ref{sec:atmosphere}, we review the physics of turbulent fluctuations in atmospheric water vapor.
In Section \ref{sec:optics}, we review the optical properties of diffraction-limited telescopes and discuss the effect of near-field beams. In Section \ref{sec:implications-for-forecasting}, we present an approximate forecast for instruments with varying apertures and atmospheric conditions.
In Section \ref{sec:discussion}, we discuss the limitations of this model, and how a full simulation of the near-field optics combined with a realistic data reduction pipeline could refine these estimates.

\section{Atmospheric fluctuations}
\label{sec:atmosphere}

\begin{table*}[]
    \centering

\begin{tabular}{lrrl}
\toprule
\textit{Survey parameters} & Symbol & Fiducial value & Description \\
\midrule
Frequency & $\nu$ & 150 GHz & Observing frequency \\
Sky fraction & $f_\text{sky}$ & 0.5 & Proportion of sky to be mapped \\
Survey time & $T_\text{survey}$ & 1\,yr & Effective integration time (5 yr at 20\% observing efficiency) \\
Elevation & $\epsilon$ & 45$^\circ$ & Observing elevation \\
Primary aperture & $D$ & $10$\,m & Diameter of the primary aperture; sets $\sigma_\text{beam} = D/4$ \\
Field of view & $\theta_\text{fov}$ & $3^\circ$ & Instantaneous field of view \\
Map white noise level & $C_\text{$\ell$, white}$ & $8 \times 10^{-8}$\,$\mu$K$_\text{CMB}^2$ & Single-frequency temperature map white noise level\\
\midrule
\textit{Atmospheric parameters} \\
\midrule
Water vapor fluctuation scale & $\sigma_{\rho, \text{base}}$ & $3 \times 10^{-5}$ kg\,m$^{-3}$ & Water vapor density fluctuation amplitude at $z = 0$ \\
Scale height & $h_0$ & 1500\,m & Vertical decay length of fluctuations; sets $z_0 = h_0/\sin\epsilon$ \\
Outer scale & $r_0$ & 500\,m & Correlation length of turbulent fluctuations \\
Wind speed & $v$ & 0.5\,deg\,s$^{-1}$ & Angular wind speed scale\\
\bottomrule
\end{tabular}
    \caption{Fiducial parameters assumed in computing the different map spectra $C_\ell = C_\text{$\ell$, atm, map} + C_\text{$\ell$, white}$ in Section~\ref{sec:implications-for-forecasting}. Atmospheric parameters are taken from either \cite{morris/etal:2022} or \cite{morris2025atacama}, while survey parameters are taken as those proposed for CMB-HD~\citep{CMB-HD-Snowmass2021}, with the exception of the primary aperture $D$, which we vary in Figure~\ref{fig:N-ell-vary-primary-size}. }
    \label{tab:parameters}
\end{table*}

Fluctuations in atmospheric emission arise when turbulent vortices mix different layers of the atmosphere. Strata at different heights have different mean temperatures and concentrations of atmospheric species (e.g. water vapor), and hence this mixing leads to random variations in those parameters: for example, a large turbulent vortex can transport a cool parcel of air from higher in the atmosphere to the ground, and vice versa. This mixing is most pronounced when there is a strong vertical gradient in some parameter: for emission at millimeter and submillimeter wavelengths, fluctuations are caused almost entirely by turbulent variations in water vapor density, as humidity decreases rapidly as a function of height \citep{morris2025atacama}.

As derived in Appendix~\ref{appendix:power-derivation}, the time-dependent power fluctuation in a single detector due to water vapor can be written as 
\begin{equation}
    \delta P(t) = g_\nu \, \delta w_\text{eff}(t)
    \label{eq:pwv-power}
\end{equation}
where $g_\nu$ is the overall coupling\footnote{Typical values of $g_\nu$ are given in Table~\ref{tab:calibration}.} for an observing band centered at frequency $\nu$ to an effective line-of-sight water vapor fluctuation given by
\begin{equation}
    \delta w_\text{eff}(t) = \iiint \delta \rho (x, y, z, t) B_n(x, y, z) dx \, dy \, dz.
\end{equation}
Here $\delta \rho (x, y, z, t)$ is a fluctuation in the spatiotemporal distribution of atmospheric water vapor scaled by the air temperature, and $B_n(x, y, z)$ is the normalized beam response. We also note that in this paper, we define the $(x,y,z)$ frame to align with the pointing: $z$ is the distance from the telescope along the line of sight, and $x$ and $y$ are physical distances transverse to it, so that $x=y=0$ is always along the centerline of the beam \citep[see Figure 2 of][]{morris/etal:2022}. 

The power fluctuation $\delta P(t)$ measured as a detector timestream is autocorrelated because the distribution of water vapor fluctuations is a spatiotemporally correlated stochastic process. Per \citet{tatarski}, the instantaneous spatial distribution of a passive additive (like water vapor) in a fully developed and perfectly turbulent (zero viscosity) flow is statistically homogeneous and isotropic, and is correlated as
\begin{equation}
\label{eqn:instantaenous-spatial-turbulent-covariance}
    \left \langle \delta \rho(x, y, z) \delta \rho(x', y', z') \right \rangle \propto M_{1/3}(r/r_0).
\end{equation}
Here $M_{1/3}$ is the Matérn function of order 1/3, $r$ is the spatial distance between points $x,y,z$ and $x',y',z'$, and $r_0$ is some characteristic correlation length (the ``outer scale" of turbulent fluctuations). Since humidity falls rapidly with altitude, we allow the variance to vary in the vertical direction while retaining horizontal homogeneity. Thus,  
\begin{equation}
\label{eqn:instantaenous-spatial-atmosphere-covariance}
    \left \langle \delta \rho(x, y, z) \delta \rho(x', y', z') \right \rangle = \sigma_{\rho}(z) \sigma_{\rho}(z') M_{1/3}(r/r_0) 
\end{equation}
where the scale of fluctuations in spatial water vapor density
\begin{equation}
\label{eqn:atm-decay}
    \sigma_{\rho}(z) = \sigma_{\rho, \text{base}} e^{-z \sin \epsilon / h_0} = \sigma_{\rho, \text{base}} e^{-z/z_0}
\end{equation}
decays exponentially as a function of height $z\sin\epsilon$, in proportion to the total concentration of water vapor. Here $\sigma_{\rho, \text{base}}$ is the standard deviation of spatial water vapor density fluctuations at $z=0$, $h_0$ is the scale height parameterizing its vertical decay, and $\epsilon$ is the elevation above the horizon (and so the decay length along the beam $z_0 = h_0 / \sin \epsilon$ implicitly depends on the elevation). We also adopt the model of frozen turbulence transported by an angular horizontal wind velocity \citep{lay:1997} so that the time evolution of the atmosphere is given by
\begin{equation}
    \rho(x, y, z, t) = \rho(x + v_x z t, y + v_y z t, z, 0).
\end{equation}

\section{Near-field optics}
\label{sec:optics}

\subsection{Gaussian beams}

The primary beam of a millimeter-wave telescope can be described by a Gaussian beam, which has an unnormalized profile
\begin{equation}
    B(x, y, z) = \exp \left ( - \frac{r^2}{2\sigma_\text{beam}(z)^2} \right )
\end{equation}
where $r = \sqrt{x^2 + y^2}$ is the transverse distance from the beam axis in the plane at a distance $z$ from the receiver along the line of sight. 
The beam width $\sigma_\text{beam}(z)$ is given by
\begin{equation}
\label{eqn:beam-width}
    \sigma_\text{beam}(z) = \sigma_\text{beam} \sqrt{1 + \left ( \frac{z}{z_R} \right )^2 }
\end{equation}
with the Rayleigh range $z_R$, the distance over which diffraction broadens the beam to $\sqrt2$ times its original width, given by
\begin{equation}
    \label{eqn:rayleigh-range}
    z_R = 
    \frac{4\pi \sigma_\text{beam}^2}{\lambda}.
\end{equation}

In the far-field limit $z \gg z_R$, the beam for a given angular radius $\theta = r/z$ becomes
\begin{equation}
    B(\theta) = \exp \left ( - \frac{\theta^2}{2\theta_\text{beam}^2} \right )
\end{equation}
where the angular beam standard deviation is given by
\begin{equation}
    \theta_\text{beam} = \frac{\lambda}{4\pi \sigma_\text{beam}}.
\end{equation}
The total angular beam width, defined as the full width at $1/e^2$, is then
\begin{equation}
    \label{eqn:total-angular-beam-size}
    \Theta_\text{beam} = 4\theta_\text{beam} =  \frac{\lambda}{\pi \sigma_\text{beam}}.
\end{equation}
For an aperture of diameter $D$, the full width at $1/e^2$ in the far field is given by the Rayleigh criterion
\begin{equation}
    \Theta_\text{beam} = 1.22\frac{\lambda}{D}.
\end{equation}
Associating these two equations gives us 
\begin{equation}
    1.22 \pi \sigma_\text{beam} = D
\end{equation}
or $\sigma_\text{beam} \approx D/4$.
In the near field $z \ll z_R$, we also see that setting the Gaussian's one-dimensional width $\sigma_\text{beam}$ equal to the
one-dimensional RMS width of a uniformly illuminated circular primary of
diameter $D$, $\langle x^2\rangle^{1/2} = D/4$ (where $x$ is a transverse
coordinate across the aperture) gives $\sigma_\text{beam} = D/4$. 
This is broadly consistent with measurements showing that the illumination of the primary extends to nearly its full diameter~\citep{Gallardo2012}.  In this regime,
Equation~\ref{eqn:beam-width} gives a roughly constant beam width $\sigma_\text{beam}(z) = \sigma_\text{beam}$. 

\subsection{Beams for high-resolution surveys}

In order to probe the sky at small angular scales, millimeter-wave telescopes must employ large apertures. Observing with a resolution of an arcminute or finer on the sky at 150 GHz necessitates a primary with a diameter $D \gtrsim 8$~m. This implies a Rayleigh range of $z_R \gtrsim 25\,$km, more than an order of magnitude above the water vapor scale height $h_0$ and above essentially all of the water vapor in the atmosphere.  The emitting column thus lies well within the near field.

The actual beam profile depends on the particular optical design of the telescope, which produces features like sidelobes that a Gaussian cannot capture; these have been studied for the Atacama Cosmology Telescope (ACT) by \citet{dunner2012far} and~\citet{gallardo2018far}.  A full treatment of the beam's effect on atmospheric noise would require measurements of the beam profile from, for example, a drone-mounted source. We emphasize, however, that any beam model (even a non-analytic one) can be substituted into the analysis in this paper, provided its Fourier transform can be computed.

\subsection{The effect of near-field beams on angular atmospheric power spectra}

From the derivation in Appendix~\ref{appendix:atmosphere-spectrum}, which re-expresses the vertically correlated turbulent field as a stack of independent layers so that the distance-dependent beam can be applied to each, we can approximate the instantaneous atmospheric angular power spectrum as
\begin{multline}
    C_\text{$\ell$, atm, inst} = 8\pi^{3/2} \frac{\Gamma(\frac{11}{6}) }{\Gamma(\frac{1}{3})} \frac{g_\nu^2 \sigma_{\rho, \text{base}}^2}{r_0^{2/3}}  \ell^{-11/3} \\ \times \int dz \, z^{5/3} \exp \left [ -\frac{\sigma_\text{beam}(z)^2 \ell^2}{z^2} - \frac{2z}{z_0} \right ].
    \label{eq:inst-power}
\end{multline}
This is the angular power spectrum of an instantaneous snapshot of the atmospheric emission as seen by the instrument.
The first term in the exponential is the effect of the beam, which suppresses structure on angular scales smaller than the footprint $\sigma_\text{beam}(z)$ subtends at a distance $z$, i.e., $\ell \gtrsim z/\sigma_\text{beam}(z)$. Since in the near field that footprint is a fixed physical size rather than growing with distance as it would in the far field, the suppression is strongest for the atmosphere nearest the telescope and stronger for larger apertures. The second term is the decline of the water vapor fluctuations with distance along the line of sight.

When many scans over the atmosphere are binned\footnote{We assume in this work that samples are binned directly into the map without any filtering or weighting. This is a very suboptimal way to make a map, and most millimeter-wave telescopes use a maximum likelihood algorithm as part of their data reduction, which reduces noise by weighting different Fourier modes in the time-ordered data. The small angular scales considered in this work are usually white noise dominated in the time domain, though, meaning that a maximum likelihood mapper weighs all modes equally and thus essentially bins the data. 
} into a map, the spectrum becomes
\begin{multline}
\label{eqn:atmosphere-map-power-spectrum}
    C_\text{$\ell$, atm, map} = 
    128 \sqrt{2} \pi^{2} \frac{\Gamma(\tfrac{11}{6})}{\Gamma(\tfrac{1}{3})} \frac{f_\text{sky}}{\theta_\text{fov}^2 T_\text{survey}} \frac{g_\nu^2 \sigma_{\rho, \text{base}}^2}{vr_0^{2/3}}  \ell^{-14/3} \\ \times \int dz \, z^{5/3}  \exp \left [ -\frac{\sigma_\text{beam}(z)^2 \ell^2}{z^2} - \frac{2z}{z_0} \right ],
\end{multline}
where we assume a frozen-flow advection of the turbulent atmosphere past the beam, a single characteristic angular wind speed independent of height, a uniformly distributed direction across all scans, and the flat-sky limit $C_\ell = P(\ell)$, as detailed in Appendices~\ref{appendix:atmosphere-spectrum-map} and~\ref{appendix:k-to-ell}. Descriptions of the input parameters along with their fiducial values are described in Table~\ref{tab:parameters}. The map spectrum is steeper than the instantaneous one by one power of $\ell$, as structure on smaller scales is averaged down more quickly by the wind modulating the atmosphere within the beam.\footnote{Empirical noise spectra from maximum likelihood maps fall roughly as $\ell^{-4}$ at these multipoles
\citep[see e.g. Figure 35 of][]{naess2025atacama}, comparable to the $\ell^{-14/3}$ power law, though a direct comparison with our model would require accounting for the mapmaker transfer function, the range of observing conditions, non-atmospheric noise sources, etc.}
Equations~\ref{eq:inst-power} and~\ref{eqn:atmosphere-map-power-spectrum} hold in both near- and far-field regimes; 
$\sigma_\text{beam}(z)$ is given by Equation~\ref{eqn:beam-width} at any distance, and in the far field, $\sigma_\text{beam}(z)/z \to \theta_\text{beam}$ recovers a constant angular beam.

\section{Predicted noise spectra}
\label{sec:implications-for-forecasting}

\begin{figure}[t]
\includegraphics[width=0.5\textwidth]{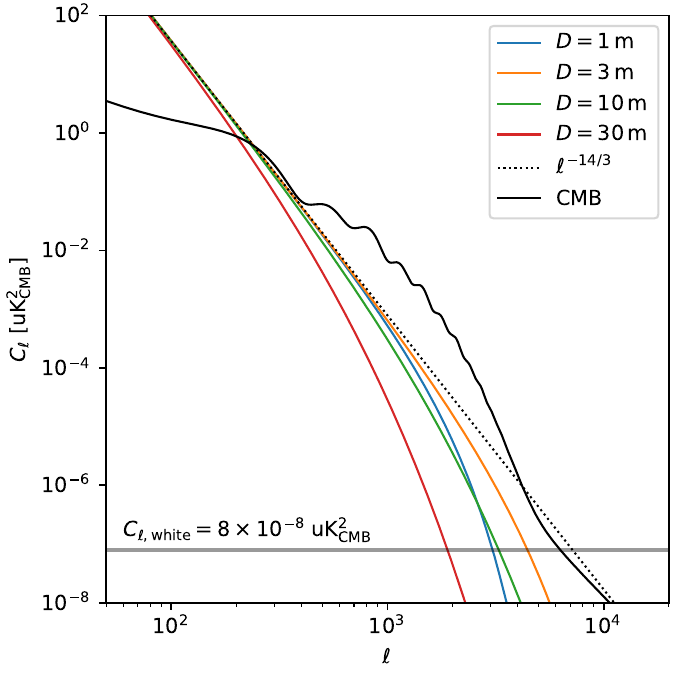}
\caption{Modeled atmospheric map noise spectra from our toy model, computed from Equation~\ref{eqn:atmosphere-map-power-spectrum}, for different primary diameters $D$. All other parameters take the value specified in Table~\ref{tab:parameters}. The black solid curve is the lensed CMB temperature power spectrum, and the grey horizontal line indicates the white noise level at 150 GHz for a fiducial 5-year survey.  The dotted line shows the standard power-law approximation for the atmosphere, normalized to the $D=10$\,m curve at $\ell = 100$, where beam effects are negligible. We see that for large apertures, the atmospheric noise can be significantly lower than the standard power-law assumption.}
\label{fig:N-ell-vary-primary-size}
\end{figure}

\begin{figure}[t]
\includegraphics[width=0.5\textwidth]{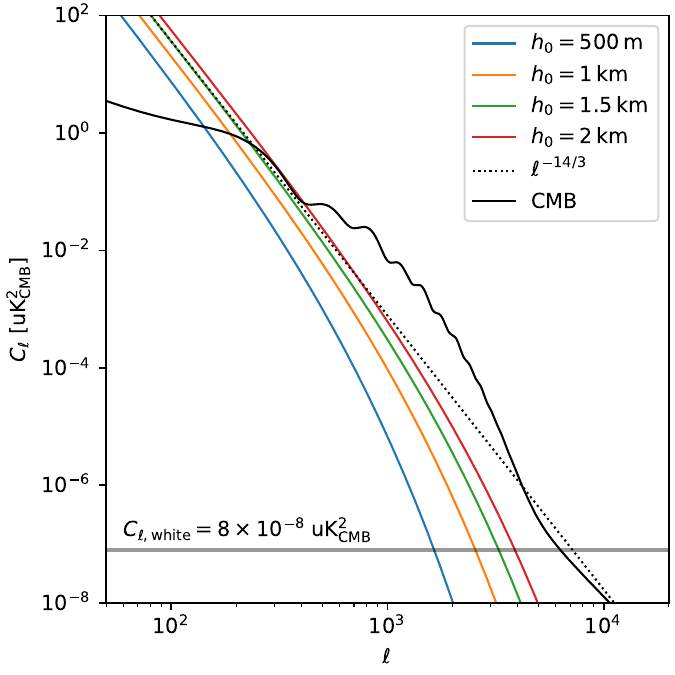}
\caption{Same as Figure~\ref{fig:N-ell-vary-primary-size}, but fixing the primary aperture $D=10$\,m and varying the scale height $h_0$, which parameterizes the vertical decay length of atmospheric fluctuations; $\sigma_{\rho,\text{base}}$ is rescaled to keep $\langle\delta w^2\rangle$ fixed (see Section~\ref{sec:atm-params}). The dotted line is the power-law approximation, normalized to the fiducial curve at $\ell = 100$. We see that increasing the scale height raises the atmospheric spectrum, most strongly at high $\ell$. }
\label{fig:N-ell-vary-scale-height}
\end{figure}

\begin{figure}[t]
\includegraphics[width=0.5\textwidth]{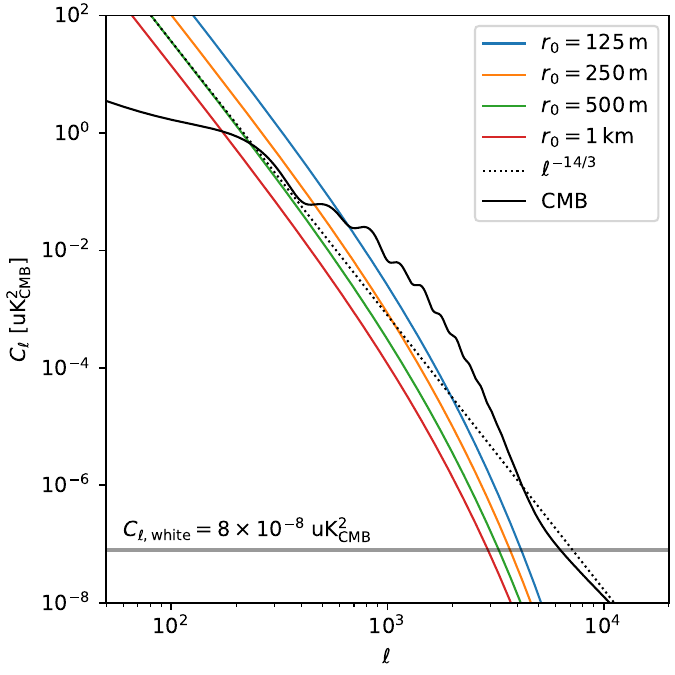}
\caption{Same as Figure~\ref{fig:N-ell-vary-scale-height}, but varying the outer scale $r_0$, which is the characteristic correlation length of turbulent fluctuations. Increasing $r_0$ lowers the atmospheric spectrum by an overall factor without changing its shape.}
\label{fig:N-ell-vary-outer-scale}
\end{figure}

Since this study is more relevant for deeper maps, we adopt
$C_{\ell,\text{white}} = {8\times10^{-8}}\,\mu\text{K}^2_\text{CMB}$ as a fiducial
white noise level, which is the level expected for 150 GHz temperature maps from the
proposed CMB-HD experiment \citep{sehgal2019cmb, CMB-HD-Snowmass2021} over a
5-year survey with 20\% observing efficiency.

\subsection{The effect of survey parameters}

The computed map spectrum depends heavily on the primary aperture of the telescope: the near-field beam has the effect of suppressing atmospheric noise at small angular scales, with the effect being more pronounced for larger $\ell$, and, for apertures above a few meters, for larger apertures, as seen in Figure~\ref{fig:N-ell-vary-primary-size}. 
Specifically, given Equations~\ref{eqn:beam-width} and~\ref{eqn:rayleigh-range}, we see that the beam width at a given distance $z$ is minimized when $\sigma_\text{beam} = \sqrt{\lambda z/4\pi}$.
For $\lambda$ = 2\,mm (150\,GHz) and $z=1$\,km, this gives $D = 4\sigma_\text{beam} \sim 1.6$\,m. 
This is the aperture that averages over the smallest physical area of the turbulent layer and thus suffers the most atmospheric noise from that distance.
Apertures well below this suppress the atmosphere because their beams have diverged before reaching the turbulent layer, and apertures well above this suppress the atmosphere because they illuminate a large physical area of it.  We see this behavior in Figure~\ref{fig:N-ell-vary-primary-size}, which shows atmospheric noise curves as a function of primary aperture size.
For apertures well within the near-field regime, the atmospheric suppression is also independent of frequency since $\sigma_\text{beam} = D/4$ contains no wavelength. The frequency dependence in Equation~\ref{eqn:atmosphere-map-power-spectrum} only changes the amplitude and not the shape of the atmospheric noise.
The field of view also affects the final map spectrum, entering Equation~\ref{eqn:atmosphere-map-power-spectrum} as $\theta_\text{fov}^{-2}$. Because the atmosphere is so heavily correlated, it averages down much more quickly (given the same total number of samples) when sampled sparsely over many independent scans (a large field of view) than densely over a few scans (a small field of view). 
We also note that, for our model, the multipole where the atmospheric noise crosses the white noise (called the atmospheric knee) is set by the instrument and observing site and not the survey plan.  In particular, it is set by the primary aperture, field of view, and detector array sensitivity, together with the properties of the turbulent layer and wind speed, and not by the sky fraction or survey duration, since those scale the atmospheric and instrument noise identically. Figure~\ref{fig:N-ell-vary-primary-size} shows that the atmospheric knee for large primary apertures occurs at a much lower multipole than expected from a naive power-law extrapolation (indicated by the dotted curve).

\subsection{The effect of atmospheric parameters}
\label{sec:atm-params}

While we explicitly parameterize the total scaling of spatial water vapor density fluctuations with $\sigma_{\rho, \text{base}}$, we obtain its fiducial value implicitly for this paper by calibrating to the magnitude of fluctuations in the total precipitable water vapor $w$. The total water vapor at the zenith is simply the integral of water vapor density at all heights
\begin{equation}
    w = \int_0^\infty \rho(h) dh,
\end{equation}
but the variance of $w$ is more complicated because fluctuations in $\rho(h)$ are correlated between different heights $h$. We therefore have 
\begin{equation}
    \langle \delta w^2 \rangle = \left \langle \iint \delta \rho(h) \delta \rho(h') \, dh \, dh' \right \rangle
\end{equation}
\begin{equation}
    = \iint \left \langle \delta \rho(h) \delta \rho(h')  \right \rangle \, dh \, dh'
\end{equation}
\begin{equation}
    = \sigma_{\rho, \text{base}}^2 \iint e^{-(h + h')/h_0} M_{1/3} \left (|h'-h|/r_0 \right ) \, dh \, dh'
\end{equation}
substituting Equations~\ref{eqn:instantaenous-spatial-atmosphere-covariance} and~\ref{eqn:atm-decay}; the magnitude of the total water vapor fluctuation depends on the scale height $h_0$ and the outer scale $r_0$.
While typical magnitudes change both seasonally and diurnally, we fix $\langle \delta w^2 \rangle = (20$\,$\mu\text{m})^2$ as a fiducial value for Chajnantor \citep[see Figure 2 in ][]{morris2025atacama}. We define $\sigma_{\rho, \text{base}}$ as whatever value recovers this value for a given $h_0$ and $r_0$, meaning that varying either $h_0$ or $r_0$ also implicitly varies $\sigma_{\rho, \text{base}}^2$.

We present the results of varying $h_0$ and $r_0$ in Figures~\ref{fig:N-ell-vary-scale-height} and~\ref{fig:N-ell-vary-outer-scale}.  From Figure~\ref{fig:N-ell-vary-scale-height}, we see that increasing the scale height $h_0$ raises the atmospheric spectrum, most strongly at high $\ell$.  Two effects lead to this: turbulence at a greater distance projects onto smaller angular scales, and in the near field, the beam's angular footprint $\sigma_\text{beam}/z$ is largest close to the telescope, so nearby water vapor is averaged away more effectively. A larger $h_0$ puts more of the water vapor at greater distances, where this suppression is weaker.  

Increasing the outer scale $r_0$, as shown in Figure~\ref{fig:N-ell-vary-outer-scale}, lowers the atmospheric spectrum by an overall factor, as reflected in the $\sigma^2_{\rho, \text{base}} r_0^{-2/3}$ prefactor of Equation~\ref{eqn:atmosphere-map-power-spectrum}. This is because the total water vapor fluctuation, which is held fixed, is spread among turbulent eddies of all sizes up to $r_0$. A larger outer scale thus leaves less fluctuation on the small scales probed here. For both $h_0$ and $r_0$, since the spectra fall steeply near the knee, even large changes in the atmospheric power shift the atmospheric knee only modestly.

\section{Discussion and Conclusions}
\label{sec:discussion}

We have presented a semianalytic toy model that shows the appreciable effect of near-field optics in suppressing atmospheric noise on small angular scales and derived its impact on the map-level spectrum to show how it scales with aperture size and atmospheric parameters. Since a large-aperture telescope's beam remains roughly the width of its primary out to distances well above the water vapor layer, the telescope averages over turbulent structures close to it and cannot resolve them on small scales. Forecasting noise spectra therefore requires a treatment of the whole three-dimensional beam and turbulent atmosphere in order to accurately predict noise at small angular scales. 

We find that the near-field beam suppresses atmospheric noise well below the standard power-law expectation at high multipoles, by orders of magnitude for apertures of tens of meters.  This suppression is weakest for apertures of a few meters and grows steadily for larger apertures.  The resulting atmospheric knee is set by the instrument and site, not by the sky fraction or survey duration.  
Our computed atmospheric spectra are overestimates on large angular scales where the time-ordered data are dominated by atmospheric noise ($\ell \lesssim 400$ for a $\sim$1\,deg/s scan and a $\sim$1\,Hz atmospheric knee per detector in the time domain),
since maximum likelihood mapmaking down-weights these modes.  However, they should be more representative of the spectrum on smaller scales, where the time-ordered data are dominated by white noise.

We caution against using this work directly as a forecast for any given telescope, as our analysis pertains only to the effects of the atmosphere and not the many other systematics relevant to millimeter-wave telescopes. Ground pickup, thermal effects from scans or cryostats, detector glitches and drifts, and other instrument-related systematics contribute to the noise spectrum at all angular scales. This work is also limited by the simplifications made in the survey design in order to achieve a more analytical result; in reality, scans are not uniform on the sky (in position or direction) and would couple strongly to the diurnal and seasonal variation in the atmospheric conditions. A rigorous forecast would combine a full simulation of the near-field optics, including measured beam profiles, with a realistic atmospheric model, instrument systematics, and maximum likelihood mapmaking. 
Nonetheless, our results show that atmospheric noise need not limit the small-scale sensitivity of large-aperture millimeter-wave telescopes.\\

\section*{Acknowledgments}
\label{sec:acknowledgments}

TM and NS thank Amanda MacInnis, Tony Mroczkowski, Sigurd Naess, Laura Newburgh, and Lyman Page for useful discussions. NS acknowledges support from DOE award number DE-SC0025309. The code used to derive and compute the spectra in this paper is available at \url{https://github.com/thomaswmorris/near-field-beam-spectra}.

\raggedbottom

\bibliographystyle{act_titles_sigurd}
\bibliography{refs.bib}

\appendix

\setcounter{table}{0}
\renewcommand{\thetable}{A\arabic{table}}

\setcounter{figure}{0}
\renewcommand{\thefigure}{A\arabic{figure}}

\section{Incident detector powers}
\label{appendix:power-derivation}

\subsection{Total detector power}

Assuming that the atmosphere is optically thin (i.e., that no emission is reabsorbed on its way to the receiver), the time-dependent power incident on a detector from a single polarized mode is
\begin{equation}
    P(t) = \frac{1}{2} \iiiint j_\nu(\phi, \theta, z, t) p(\nu) \, d\nu \, dA \, d\Omega \, dz,
\end{equation}
where we integrate the emission $j_\nu$ weighted by the instrument passband $p(\nu)$ (which includes an overall efficiency) over all volume elements in the atmosphere $d\Omega \, dz$, the collecting area of the telescope $dA$, and the frequency spectrum $d\nu$.
In the Rayleigh-Jeans limit, we have for each volume element of the atmosphere at temperature $T$ and frequency $\nu$ (and associated wavelength $\lambda$)
\begin{equation}
    j_\nu = \frac{2k_\text{B}T}{\lambda^2} \kappa(\nu)
\end{equation}
where $T$ is its physical temperature, $\lambda$ is the associated wavelength, and $\kappa(\nu)$ is its absorption coefficient. An ideal radio antenna has an effective collecting area\footnote{See e.g. Section 3.1.4 of \cite{condon/ransom:2016}.}
\begin{equation}
    A_e = \frac{\lambda^2 B(\phi, \theta)}{\Omega_\text{beam}}
\end{equation}
where $B$ describes the beam at each point on the sphere and $\Omega_\text{beam}$ is its total solid angle; we neglect the beam's frequency dependence because we assume that the passband is over a narrow enough frequency range such that the shape of the beam can be assumed to be constant.
We can then write
\begin{equation}
    P(t) = k_\text{B} \iiint T(\phi, \theta, z, t) \kappa(\nu, \phi, \theta, z, t) p(\nu)\frac{B(\phi, \theta)}{\Omega_\text{beam}} \, d\nu \,  d\Omega \, dz
\end{equation}
Reformulating our integral as an integral over $dx\,dy\,dz = z^2 d\Omega \, dz$, we have
\begin{equation}
    P(t) = k_\text{B} \iiint T(x, y, z, t) \kappa(\nu, x, y, z, t) p(\nu)B_n(x, y, z) \, d\nu \,  dx \, dy \, dz
\end{equation}
where
\begin{equation}
    B_n(x, y, z) = \frac{B(x, y, z)}{z^2 \Omega_\text{beam}}
\end{equation}
such that $\iint B_n(x, y, z) \, dx \, dy = 1$. The temperature of the atmosphere decreases as a function of height, but much more slowly in relative terms than the density of water vapor; thus, we assume that the temperature of the atmosphere is a constant $T_\text{atm}$.\footnote{The effect of such a temperature decrease would be absorbed into the scale height governing the decay of atmospheric fluctuations.} We assume the absorption coefficient of water is proportional at each frequency to the spatial density $\rho(x, y, z)$ of water in the atmosphere, and so we have
\begin{equation}
    \kappa(\nu, x, y, z, t) = a(\nu) \rho(x, y, z, t)
\end{equation}
where $a(\nu)$ is a proportionality coefficient, and so we can now write
\begin{equation}
    P = g_\nu \iiint \rho(x, y, z, t) B_n(x, y, z) \,  dx \, dy \, dz
    \label{eq:pwv-power-2}
\end{equation}
where 
\begin{equation}
    g_\nu = k_\text{B} T_\text{atm} \int a(\nu') p(\nu') \, d\nu'
\end{equation}
is the power received in a band centered at $\nu$ per unit of water vapor along the line of sight.

\subsection{Calibrating fluctuations to the CMB}

To compute the instrument response to water vapor $g_\nu$, we integrate the detector passband over the atmospheric spectrum computed from \texttt{am} \citep{am:2018} and take the partial derivative of incident power $P$ with respect to the line-of-sight water vapor $w_\text{eff}$.
We convert this to temperature units by computing the power derivative from a perfect Rayleigh-Jeans source ($T_\text{RJ}$) and a blackbody at 2.73\,K ($T_\text{CMB}$). We can then compute
\begin{equation}
    g_\nu = \frac{\partial T_\text{CMB}}{\partial w_\text{eff}} = \frac{\partial P}{\partial w_\text{eff}} \left ( \frac{\partial P}{\partial T_\text{CMB}} \right )^{-1}
\end{equation}
for a given passband centered at $\nu$. We present typical values of $g_\nu$ in the last row of Table~\ref{tab:calibration} for the six frequency bands of ACT \citep{alford2026atacama}. 

\begin{table}[]
    \centering
\begin{tabular}{lrrrrrr}
\toprule
 & 30 GHz & 40 GHz & 90 GHz & 150 GHz & 220 GHz & 280 GHz \\
\midrule
$\partial T_\text{CMB} / \partial T_\text{RJ}$ (K$_\text{CMB}$/K$_\text{RJ}$) & 1.02 & 1.04 & 1.25 & 1.70 & 3.23 & 5.85 \\
$\partial P / \partial T_\text{RJ}$ (pW/K$_\text{RJ}$) & 0.06 & 0.24 & 0.47 & 0.48 & 0.76 & 0.76 \\
$\partial P / \partial T_\text{CMB}$ (pW/K$_\text{CMB}$) & 0.06 & 0.23 & 0.38 & 0.28 & 0.24 & 0.13 \\
$\partial P / \partial w$ (pW/mm) & 0.04 & 0.08 & 0.76 & 2.45 & 8.93 & 14.05 \\
$\partial T_\text{RJ} / \partial w$ (K$_\text{RJ}$/mm) & 0.60 & 0.35 & 1.61 & 5.08 & 11.73 & 18.39 \\
$\partial T_\text{CMB} / \partial w$ (K$_\text{CMB}$/mm) & 0.61 & 0.37 & 2.01 & 8.62 & 37.86 & 107.55 \\
\bottomrule
\end{tabular}
    \caption{Several calibration factors using ACT passbands \citep{alford2026atacama}, computed with \texttt{am} for observations at the zenith. The last three rows assume fluctuations around a mean zenith water vapor of $w = 1$\,mm. The different values of $g_\nu$ for each band are in the last row.}
    \label{tab:calibration}
\end{table}

\section{Scans and indicator functions}

Consider a circular map of diameter $D_\text{map}$ (and hence total area $\Omega_\text{map} = \pi D_\text{map}^2/4$) with pixels of solid angle $\Omega_\text{pix}$, and an array of $N_\text{det}$ detectors distributed uniformly over a square focal plane of side length $\theta_\text{fov}$ (with total area $\Omega_\text{fov} = \theta_\text{fov}^2$). A survey consists of the array performing many ``scans" over the map at a constant speed $s$ at various angles $\phi$ and offsets $a$ such that for any given scan parameterized by $a$ and $\phi$ the boresight of the array as a function of time is
\begin{equation}
    \begin{cases}
        \theta_x^\text{boresight}(t) = s t \cos \phi - a \sin \phi \\
        \theta_y^\text{boresight}(t) = s t \sin \phi + a \cos \phi
    \end{cases}
\end{equation}
For a map of diameter $D_\text{map}$, $a$ will run from $-D_\text{map}/2$ to $D_\text{map}/2$ and $\phi$ will run from 0 to $2\pi$. If $a$ and $\phi$ are uniform, then every pixel in the circular map will be uniformly sampled and thus will have a uniform noise spectrum.
A pixel will be sampled by some detector if it is within the field of view of the array; this can be written in terms of an indicator function
\begin{equation}
    \Theta(\theta_x, \theta_y, t, a, \phi) = \Theta_t(\theta_x, \theta_y, t, \phi)\Theta_a(\theta_x, \theta_y, a, \phi)
\end{equation}
where
\begin{equation}
    \Theta_t(\theta_x, \theta_y, t, \phi) = \begin{cases}
        1 & -\frac{\theta_\text{fov}}{2} < \theta_x \cos \phi + \theta_y \sin \phi - 
        s t < \frac{\theta_\text{fov}}{2} \\
        0 & \text{otherwise} \\
    \end{cases}
\end{equation}
and 
\begin{equation}
    \Theta_a(\theta_x, \theta_y, a, \phi) = \begin{cases}
        1 & -\frac{\theta_\text{fov}}{2} < -\theta_x \sin \phi + \theta_y \cos \phi - 
        a < \frac{\theta_\text{fov}}{2} \\
        0 & \text{otherwise} \\
    \end{cases}
\end{equation}
so that $\Theta = 1$ when a pixel is being sampled by the array, and 0 when it is not. When a pixel is being sampled, we assume it is sampled at an average uniform rate $f_\text{array}$ given by
\begin{equation}
    f_\text{array} = f_\text{det}  N_\text{det}\frac{\Omega_\text{pix}}{\theta_\text{fov}^2}  
\end{equation}
where $f_\text{det}$ is the sample rate per detector in Hz. For multiple scans, the total number of samples for a pixel at coordinates $\theta_x, \theta_y$ is given by 
\begin{equation}
N_\text{per pixel} = f_\text{array}\,T_\text{dwell}, \qquad
T_\text{dwell} \equiv \sum_{i\in S}\int_{t=-\infty}^{t=+\infty} dt\,
\Theta(t,\theta_x,\theta_y,a_i,\phi_i),
\end{equation}
where $T_\text{dwell}$ is the total time the pixel spends in the field
of view over all scans, and 
$a_i$ and $\phi_i$ parameterize each scan $i$ in the set of scans $S$. In the limit of many scans (such that the parameters $a$ and $\phi$ are uniformly sampled), we can write the sum over scans as an integral
\begin{equation}
\label{eqn:sum-over-scans}
    \sum_{i \in S} \equiv N_\text{scans} \frac{1}{2 \pi} \frac{1}{D_\text{map}} \int_{\phi=0}^{\phi=2\pi} \int_{a=-D_\text{map}/2}^{a=+D_\text{map}/2} d\phi \, da
\end{equation}
where $N_\text{scans}$ is the total number of scans.

\section{Map depth and white noise spectra}

Because all pixels in the map are sampled uniformly, we consider without loss of generality the pixel at the center of the map with coordinates $\theta_x, \theta_y = (0, 0)$. That pixel will be sampled if and only if $-\theta_\text{fov}/2 < a < \theta_\text{fov}/2$ and $-\theta_\text{fov}/(2s) < t <\theta_\text{fov}/(2s)$, and so that pixel will have a total number of samples
\begin{equation}
    N_\text{per pixel} = f_\text{array}\sum_{i \in S} \int_{t=-\infty}^{t=+\infty} dt \, \Theta(t, 0, 0, a_i, \phi_i) = \frac{N_\text{scans} f_\text{array} }{2 \pi D_\text{map}} \int_{a=-\theta_\text{fov}/2}^{a=+\theta_\text{fov}/2} \int_{\phi=0}^{\phi = 2 \pi} \int_{t=-\theta_\text{fov}/(2s)}^{t=+\theta_\text{fov}/(2s)} \, dt \, d \phi \, da
\end{equation}
\begin{equation}
\label{eqn:Npix}
    = \frac{f_\text{det} N_\text{det}N_\text{scans}\Omega_\text{pix}}{D_\text{map}s}. 
\end{equation}
The total time the focal plane dwells on a pixel is $T_\text{dwell} = N_\text{per pixel}/f_\text{array} = \theta_\text{fov}^2 N_\text{scans}/(D_\text{map}\,s)$,
while the total detector-time per pixel is
\begin{equation}
\label{eqn:Tpix}
    T_\text{per pixel} = \frac{N_\text{per pixel}}{f_\text{det}} = \frac{N_\text{det}N_\text{scans}\Omega_\text{pix}}{ D_\text{map}s}. 
\end{equation}
These differ by $T_\text{per pixel}/T_\text{dwell} = N_\text{det}\Omega_\text{pix}/\theta_\text{fov}^2$, the number of detectors simultaneously covering the pixel.  Since detector noise is independent between detectors, we use $T_\text{per pixel}$ to set the white noise level.\\

The time to scan across the map is $T_\text{per scan} = D_\text{map} / s$, and so during a survey of time $T_\text{survey}$ we have $N_\text{scans} = T_\text{survey} / T_\text{per scan}$. This gives us a variance per pixel
\begin{equation}
    \sigma_\text{pix}^2 = \frac{\text{NET}_\text{CMB}^2}{T_\text{per pixel}} = \frac{D_\text{map} s}  {N_\text{det}N_\text{scans}\Omega_\text{pix}}\text{NET}_\text{CMB}^2 = \frac{D_\text{map}^2}  {N_\text{det}T_\text{survey}\Omega_\text{pix}}\text{NET}_\text{CMB}^2
\end{equation}
and a map depth
\begin{equation}
    \sigma_\text{pix} \sqrt{\Omega}_\text{pix} = \sqrt{\frac{D_\text{map} s}{N_\text{det} N_\text{scans}}} \text{NET}_\text{CMB} = \sqrt{\frac{D_\text{map}^2 }{N_\text{det} T_\text{survey}}} \text{NET}_\text{CMB}  = 4\sqrt{\frac{f_\text{sky} }{T_\text{survey}}} \text{NET}_\text{array}
\end{equation}
where $\text{NET}_\text{array} = \text{NET}_\text{det} / \sqrt{N_\text{det}}$ and $f_\text{sky} = \Omega_\text{map}/4\pi = D_\text{map}^2/16$ is the fraction of the sky covered by the map. The map depth has no dependency on the packing density of the detectors, and so we can intuit that this result is valid even when detectors are not uniformly distributed on the focal plane (e.g. when they are clustered into optics tubes).

The white noise spectrum is the Fourier transform of the white noise covariance 
\begin{equation}
    P_\text{white}(k_x, k_y) = \iint C(\theta_x, \theta_y) e^{-i(k_x\theta_x + k_y \theta_y)} dx \, dy
\end{equation}
If the pixel is very small relative to $1/k$, then $e^{-i(k_x\theta_x + k_y \theta_y)} \approx 1$ and we have
\begin{equation}
     P_\text{white}(k_x, k_y) = \int_{\Omega_\text{pix}} \sigma_\text{pix}^2 d\Omega = \sigma_\text{pix}^2 \Omega_\text{pix}
\end{equation}
which is just the square of the map depth. From Appendix~\ref{appendix:k-to-ell}, the multipole white noise spectrum is likewise just $C_\text{$\ell$, white} = P_\text{white}(\ell, 0) = \sigma_\text{pix}^2 \Omega_\text{pix}$.

\section{Deriving the angular atmospheric power spectrum}
\label{appendix:atmosphere-spectrum}

\subsection{Deriving the integrated turbulent atmospheric spectrum}
\label{appendix:wind}

A detector $i$ observes the angular coordinate $(0, 0)$ at time $t_i$ and sees a temperature fluctuation
\begin{equation}
    \delta T_i(t_i) = g_\nu \iiint \delta \rho(x, y, z, t_i) B_n(x, y, z) \, dx \, dy \, dz
    \label{eq:temp-fluc-i}
\end{equation}
where we define the coordinate frame $x,y,z$ such that $x=y=0$ aligns with the centerline of the beam of detector $i$ and $z$ is in the direction that the detector is observing. In this frame, another detector $j$ observing angular coordinate $(\theta_x, \theta_y)$ at time $t_j$ will see
\begin{equation}
    \delta T_j(t_j) = g_\nu \iiint \delta \rho(x + z \theta_x, y + z \theta_y, z, t_j) B_n(x, y, z) \, dx \, dy \, dz.
    \label{eq:temp-fluc-j}
\end{equation}
This geometry follows that of~\citet{morris/etal:2022}; see their Figure 2 for a schematic of an array observing through a turbulent atmosphere and their Figure 6 for the coordinate frames of two detectors separated by an angle $\theta$. \\

The covariance of these fluctuations is the expectation value
\begin{multline}
    C(\theta_x, \theta_y, t_i, t_j) = \langle \delta T_i(t_i) \, \delta T_j(t_j) \rangle = g_\nu^2 \bigg \langle \iiint \delta \rho(x, y, z, t) B_n(x, y, z) dx \, dy \, dz \\ \times \iiint \delta \rho(x + z \theta_x, y + z \theta_y, z, t_j) B_n(x, y, z) \, dx \, dy \, dz \bigg \rangle
\end{multline}
and so we have
\begin{multline}
    C_\text{atm}(\theta_x, \theta_y, t_i, t_j) = g_\nu^2 \iiint \iiint \left \langle \delta \rho(x_i, y_i, z_i, t_i) \delta \rho (x_j + z_j \theta_x, y_j + z_j \theta_y, z_j, t_j) \right \rangle \\ \times B(x_i, y_i, z_i) B(x_j, y_j, z_j) dx_i \, dy_i \, dz_i \, dx_j \, dy_j \, dz_j
\end{multline}
The expectation in the above expression is determined by turbulent atmospheric physics.
First, we take the spatiotemporal fluctuations in water vapor density to be a frozen distribution transported horizontally by some angular wind velocity $v_x, v_y$ such that
\begin{equation}
    \rho(x, y, z, t) = \rho(x + v_x z t, y + v_y z t, z)
\end{equation}
Second, we take from \cite{tatarski} that the spatial covariance of density fluctuations for pairs of volume elements in a three-dimensional turbulence distribution is given by
\begin{equation}
    \left \langle \delta \rho(x, y, z) \delta \rho(x', y', z') \right \rangle = \sigma_{\rho}(z) \sigma_{\rho}(z') M_{1/3} \left ( \left [ (x - x')^2 + (y - y')^2 + (z - z')^2 \right ]^{1/2} / r_0 \right ) 
\end{equation}
where $\sigma_{\rho}(z) = \langle \delta \rho^2 (x, y, z) \rangle_t^{1/2} = \sigma_{\rho, \text{base}} e^{-z/z_0}$ describes the profile of fluctuation strength, and $r_0$ is the turbulent outer scale. Here $M_\mu(r) = 2^{1-\mu} r^\mu K_\mu(r) / \Gamma(\mu)$ is the normalized Matérn function, where $K_\mu$ is the modified Bessel function of the second kind of order $\mu$, and where $\mu = 1/3$ describes the spatial covariance of three-dimensional turbulence.
We therefore have
\begin{equation}
\label{eqn:C_atm_raw}
    C_\text{atm}(\theta_x, \theta_y, t_\Delta) = g_\nu^2 \sigma_{\rho, \text{base}}^2 \iiiint dx_i \, dy_i \, dx_j \, dy_j B(x_i, y_i) B(x_j, y_j) \iint dz_i \, dz_j \,\exp \left ( - \frac{z_i + z_j}{z_0} \right ) M_{1/3} \left ( \frac{d_\text{elem}}{r_0}  \right ) 
\end{equation}
where $d_\text{elem}$ is the distance between a pair of volume elements, defined as
\begin{equation}
    d_\text{elem}^2 = (x_j - x_i + v_x z (t_j - t_i) + z\theta_x)^2 + (y_j - y_i + v_y z (t_j - t_i) + z\theta_y)^2 + (z_i - z_j)^2.
\end{equation}

\subsection{Integrated turbulent atmosphere}

We first consider the $z_i$ and $z_j$ integrals (the rightmost part of Equation~\ref{eqn:C_atm_raw}). We write
\begin{equation}
    C_\text{elem}(\theta_x, \theta_y, x_i, x_j, y_i, y_j, t_i, t_j) = g_\nu^2 \sigma_{\rho, \text{base}}^2 \iint \exp \left ( - \frac{z_i + z_j}{z_0} \right ) M_{1/3} \left ( d_\text{elem} / r_0 \right ) \, dz_i \, dz_j
\end{equation}
Introducing the variables
\begin{equation}
        x = \frac{x_i + x_j}{2}, \quad
        y = \frac{y_i + y_j}{2}, \quad
        z = \frac{z_i + z_j}{2}, \quad
        x_\Delta = x_j - x_i, \quad
        y_\Delta = y_j - y_i, \quad
        z_\Delta = z_j - z_i, \quad
        t_\Delta = t_j - t_i, \quad
\end{equation}
then the above becomes
\begin{equation}
    C_\text{elem}(\theta_x, \theta_y, x_\Delta, y_\Delta, t_\Delta) = g_\nu^2 \sigma_{\rho, \text{base}}^2 \int_0^\infty dz \,\exp \left ( - \frac{2z}{z_0} \right ) \int_{-2z}^{2z} \, dz_\Delta M_{1/3} \left ( d_\text{elem} / r_0 \right ) 
\end{equation}
with
\begin{equation}
    d_\text{elem}^2 = \left (z [\theta_x + v_x t_\Delta] + x_\Delta \right )^2 + \left (z [\theta_y + v_y t_\Delta] + y_\Delta \right )^2 + z_\Delta^2.
\end{equation}
Because its integrand is even in $z_\Delta$, we can write 
\begin{equation}
    \int_{-2z}^{2z} \, dz_\Delta M_{1/3} \left ( d_\text{elem} / r_0 \right ) \approx  \int_{-\infty}^\infty dz_
    \Delta \, M_{1/3} \left ( d_\text{elem} / r_0 \right ) - 2A
\end{equation}
where 
\begin{equation}
    A = \int_{2z}^{\infty} M_{1/3} \left ( d_\text{elem} / r_0 \right ) \, dz_\Delta 
\end{equation}
is approximately constant in $\theta_x$ and $\theta_y$, because in that regime the argument in the co-element distance is dominated by the separation in $z$. Given that the mean is almost always subtracted from time-ordered data before mapping, a constant offset will not be manifest in the map noise spectrum (especially not on small scales) and hence we can ignore it and write
\begin{equation}
    C_\text{elem}(\theta_x, \theta_y, x_\Delta, y_\Delta, t_\Delta) = g_\nu^2 \sigma_{\rho, \text{base}}^2  \int_0^\infty dz \,\exp \left ( - \frac{2z}{z_0} \right ) \int_{-\infty}^{\infty} \, dz_\Delta M_{1/3} \left ( d_\text{elem} / r_0 \right ). 
\end{equation}
If we define $\theta_\text{eff}$ such that
\begin{equation}
    z^2\theta_\text{eff}^2 = d_\text{elem}^2 - z_\Delta^2 = \left (z [\theta_x + v_x t_\Delta] + x_\Delta \right )^2 + \left (z [\theta_y + v_y t_\Delta] + y_\Delta \right )^2
\end{equation}
then we can write 
\begin{equation}
    \int_{-\infty}^\infty dz_
    \Delta \, M_{1/3} \left ( d_\text{elem} / r_0 \right ) = \int_{-\infty}^\infty dz_\Delta \, M_{1/3} \left [ (z^2\theta_\text{eff}^2 + z_\Delta^2)^{1/2} / r_0 \right ]. 
\end{equation}
The above has a Fourier transform with respect to $\theta_\text{eff}$ of~\footnote{Here we use the Fourier transform identity 
\begin{equation}
    \int_{-\infty}^\infty M_{ \nu} \left [a \left (x^2 + b^2 \right )^{1/2}] \right ] e^{-i k x} dx 
    \\ = \frac{ \pi^{1/2} a^{ 2\nu} b^{2\nu + 1}}{2^{2\nu}} \frac{\Gamma(- \nu - \tfrac{1}{2})}{\Gamma( \nu)} M_{- \nu - 1/2} \left [b (k^2 + a^2)^{1/2} \right ] .
\end{equation}}
\begin{equation}
    \F{\int_{-\infty}^\infty dz_
    \Delta \, M_{1/3} \left ( d_\text{elem} / r_0 \right )} = 2^{-2/3} \pi^{1/2} \frac{\Gamma(-\frac{5}{6})}{\Gamma(\frac{1}{3})}  \int_{-\infty}^{\infty} dz_\Delta \, \left ( \frac{z_\Delta}{z} \right )^{5/3} \left ( \frac{z}{r_0} \right )^{2/3}M_{-5/6} \left (\frac{z_\Delta}{z} \left [\tilde \theta_\text{eff}^2 + \frac{z^2}{r_0^2} \right ]^{1/2} \right ),
\end{equation}
where $\tilde \theta_\text{eff}$ is the Fourier frequency corresponding to $ \theta_\text{eff}$. Carrying out the $z_\Delta$ integral gives us\footnote{
Here we use the integral identity
\begin{equation}
    \int_0^\infty x^\nu M_\mu (a x) dx = 2^\nu a^{-\nu-1} \frac{\Gamma(\mu + \tfrac{\nu+1}{2})\Gamma(\tfrac{\nu+1}{2})}{\Gamma(\mu)}
\end{equation}
}
\begin{equation}
    \F{\int_{-\infty}^\infty dz_
    \Delta \, M_{1/3} \left ( d_\text{elem} / r_0 \right )} = \frac{4\pi}{3} \frac{z^{5/3}}{r_0^{2/3}}  \left ( \tilde\theta_\text{eff}^2 + \frac{z^2}{r_0^2} \right )^{-4/3}
\end{equation}
and applying an inverse Fourier transform with respect to $\tilde \theta_\text{eff}$ gives us\footnote{
Here we use the identity 
\begin{equation}
    \int_{-\infty}^{\infty} (x^2 + b^2)^{-\nu-1/2} e^{i k x} dx \\ = \pi^{1/2} \frac{\Gamma(\nu)}{\Gamma(\nu+\tfrac{1}{2})} b^{-2\nu} M_\nu(bk)
\end{equation}
}
\begin{equation}
    \int_{-\infty}^\infty dz_
    \Delta \, M_{1/3} \left ( d_\text{elem} / r_0 \right ) = 2 \pi^{1/2} \frac{\Gamma(\tfrac{5}{6})}{\Gamma(\tfrac{1}{3})} r_0 M_{5/6} \left ( \frac{d_\text{elem}}{r_0} \right ).
\end{equation}
Our full covariance element is then
\begin{equation}
\label{eqn:final-atmosphere-covariance-element}
    C_\text{elem}(\theta_x, \theta_y, t_\Delta) = 2 \pi^{1/2} \frac{\Gamma(\tfrac{5}{6})}{\Gamma(\tfrac{1}{3})} g_\nu^2 \sigma_{\rho, \text{base}}^2  r_0  \int dz \, \exp \left ( - \frac{2z}{z_0} \right ) M_{5/6} \left ( \frac{d_\text{elem}(\theta_x, \theta_y, z, t_\Delta)}{r_0} \right )
\end{equation}
meaning that instead of summing over every correlated pair of layers of the atmosphere, we can recover the same covariance by instead summing over a set of independent layers with a different projected angular spectrum.

\subsection{Integrating over beams}

\subsubsection{Beam profiles}

We adopt a normalized beam profile
\begin{equation}
    B_n(x, y, z) = \frac{1}{2\pi \sigma_\text{beam}(z)^2} \exp \left [ -\frac{x^2 + y^2}{2\sigma_\text{beam}(z)^2} \right ],
\end{equation}
or in angular coordinates 
\begin{equation}
    B_n(\theta_x, \theta_y, z) = \frac{z^2}{2 \pi \sigma_\text{beam}(z)^2} \exp \left [ - \frac{z^2}{2\sigma_\text{beam}(z)^2} (\theta_x^2 + \theta_y^2) \right ], 
\end{equation}
normalized in either frame such that $\iint B_n \, dx\, dy = 1$. The angular version has Fourier transform\footnote{Note that in the limit $b \to 0$, we have $\tilde B_n(k_x, k_y) = 1$.}
\begin{equation}
    \tilde B_n(k_x, k_y) = \exp \left [ - \frac{\sigma_\text{beam}(z)^2}{2z^2} (k_x^2 + k_y^2) \right ], 
\end{equation}
and power spectrum
\begin{equation}
    P_\text{beam}(k_x, k_y) = \exp \left [ - \frac{\sigma_\text{beam}(z)^2}{z^2} (k_x^2 + k_y^2) \right ]. 
\end{equation} 

\subsubsection{Beam convolution}

We cannot easily perform the explicit convolution of the angular beam profiles for every pair of correlated atmospheric layers $i$ and $j$ as expressed in Equation~\ref{eqn:C_atm_raw}. However, the result of the previous section allows us to represent the atmosphere as a set of uncorrelated layers, each of which we can convolve with the beam separately (because the detectors have identical beams at the same distance $z$).

This convolution is more easily done in Fourier space, as the power spectrum of the convolution of two functions is the product of their individual power spectra. The angular power spectrum of the atmosphere is just the Fourier transform of the covariance, and so we have
\begin{equation}
    \F {C_\text{atm}} =  8 \pi^{3/2} \frac{\Gamma(\tfrac{11}{6})}{\Gamma(\tfrac{1}{3})} \frac{g_\nu^2 \sigma_{\rho, \text{base}}^2}{r_0^{2/3}} \int dz \, z^{5/3} \exp \left (-\frac{\sigma_\text{beam}(z)^2k^2}{z^2} - i t_\Delta \left [ k_x v_x + k_y v_y \right ]  - \frac{2z}{z_0} \right ) \left ( k_x^2 + k_y^2 + \frac{z^2}{r_0^2} \right )^{-11/6}.
\end{equation}

Since we care about small angular scales where $k^2 = k_x^2 + k_y^2 \gg z^2/r_0^2$ (in practice, scales with $\ell \gtrsim 100$ for our fiducial parameters), we can neglect the $z^2/r_0^2$ term, giving
\begin{equation}
\label{eqn:atmosphere-power-spectrum-element}
    P(k_x, k_y,t_\Delta) = \F {C_\text{atm}} =  8 \pi^{3/2} \frac{\Gamma(\tfrac{11}{6})}{\Gamma(\tfrac{1}{3})} \frac{g_\nu^2 \sigma_{\rho, \text{base}}^2}{r_0^{2/3}} k^{-11/3} \int dz \, z^{5/3} \exp \left (-\frac{\sigma_\text{beam}(z)^2}{z^2} k^2 - i t_\Delta \left [ k_x v_x + k_y v_y \right ]  - \frac{2z}{z_0} \right ).
\end{equation}

\section{Map noise spectra}
\label{appendix:atmosphere-spectrum-map}

\subsection{Atmospheric spectra binned in a map}

The covariance of the averages of two sets of samples is the average covariance of each pair of samples from either set.
Since we assume the atmosphere is independent between different scans, only scans that sample both pixel $i$ and pixel $j$ will contribute to their covariance in the binned map. The pixel-pixel covariance is then a sum over scans
\begin{equation}
     C_\text{map}(\theta_x, \theta_y) = 
     \frac{1}{N_\text{per pixel}^2} \sum_{i \in S}
       \iint dt_i \, dt_j \Theta(t_i, 0, 0, a, \phi) \Theta(t_j, \theta_x, \theta_y, a, \phi) f_\text{array}^2C_\text{atm}(\theta_x, \theta_y, t_j - t_i).
\end{equation}
We assume that separations $\theta = \sqrt{\theta_x^2 + \theta_y^2}$ are small enough such that when the array samples one, it always samples the other.\footnote{For separations $\theta < \theta_\text{fov}$, the fraction of samples of one pixel during which the other is also in the field of view is approximately $1 - \frac{\theta}{2\theta_\text{fov}}$, which is very close to 1 when $\theta \ll \theta_\text{fov}$.} This allows us to write 
\begin{equation}
     C_\text{map}(\theta_x, \theta_y) = 
     \frac{f_\text{array}^2}{N_\text{per pixel}^2} \sum_{i \in S}
       \iint dt_i \, dt_j \Theta(t_i, 0, 0, a, \phi) \Theta(t_j, 0, 0, a, \phi) C_\text{atm}(\theta_x, \theta_y, t_j - t_i).
\end{equation}
Note that $f_\text{array} / N_\text{per pixel} = D_\text{map}\,s / (N_\text{scans}\theta_\text{fov}^2)$ is invariant in the number of detectors and the detector sample rate, meaning that the atmosphere only averages down as the result of averaging many independent scans.
Using the integral form of the sum over scans from Equation~\ref{eqn:sum-over-scans}, we have
\begin{multline}
     C_\text{map}(\theta_x, \theta_y) = 
     \frac{f_\text{array}^2}{N_\text{per pixel}^2} \frac{N_\text{scans} }{2 \pi D_\text{map}} \\ \times \int_{\phi=0}^{\phi=2\pi} \int_{a=-\theta_\text{fov}/2}^{a=+\theta_\text{fov}/2} \int_{t_i=-\theta_\text{fov}/(2s)}^{t_i=+\theta_\text{fov}/(2s)} \int_{t_j=-\theta_\text{fov}/(2s)}^{t_j=+\theta_\text{fov}/(2s)} 
       d\phi \, da \, dt_i \, dt_j \Theta(t_i, 0, 0, a, \phi) \Theta(t_j, 0, 0, a, \phi) C_\text{atm}(\theta_x, \theta_y, t_j - t_i)
\end{multline}
\begin{equation}
      = 
     \frac{f_\text{array}^2\theta_\text{fov} N_\text{scans}}{N_\text{per pixel}^2D_\text{map}} 
       \int_{t_i=-\theta_\text{fov}/(2s)}^{t_i=+\theta_\text{fov}/(2s)} \int_{t_j=-\theta_\text{fov}/(2s)}^{t_j=+\theta_\text{fov}/(2s)} dt_i \, dt_j \, C_\text{atm}(\theta_x, \theta_y, t_j - t_i).
\end{equation}
Taking the Fourier transform with respect to $\theta_x$ and $\theta_y$, we get the power spectrum of the map
\begin{equation}
\label{eqn:sum-over-general-power-spectrum}
     P_\text{map}(k_x, k_y) = 
     \frac{f_\text{array}^2\theta_\text{fov} N_\text{scans}}{N_\text{per pixel}^2 D_\text{map}} \int_{t_i=-\theta_\text{fov}/(2s)}^{t_i=+\theta_\text{fov}/(2s)} \int_{t_j=-\theta_\text{fov}/(2s)}^{t_j=+\theta_\text{fov}/(2s)} dt_i \, dt_j \, \F{C_\text{atm}(\theta_x, \theta_y, t_j - t_i)}.
\end{equation}
\\

\subsection{Velocity-marginalized atmospheric spectra}

Because we integrate over many scans in random directions, we assume the angular wind vector $(v_x, v_y)$ is drawn randomly from a distribution
\begin{equation}
    p(v_x, v_y) = \frac{1}{2 \pi v^2} \exp \left ( -\frac{v_x^2 + v_y^2}{2v^2} \right )
\end{equation}
with some characteristic typical wind speed $v$. Marginalizing $\F{C_\text{atm}(\theta_x, \theta_y, t_j - t_i)}$ from Equation~\ref{eqn:atmosphere-power-spectrum-element} over this distribution, we have

\begin{multline}
    \left \langle \F{C_\text{atm}(\theta_x, \theta_y, t_j - t_i)} \right \rangle  = \iint p(v_x, v_y) \F{C_\text{atm}(\theta_x, \theta_y, t_j - t_i)} \, dv_x \, dv_y \\ =
    8 \pi^{3/2} \frac{\Gamma(\tfrac{11}{6})}{\Gamma(\tfrac{1}{3})} \frac{g_\nu^2 \sigma_{\rho, \text{base}}^2}{r_0^{2/3}} k^{-11/3} e^{-v^2 (t_j - t_i)^2 k^2/2 }\int dz \, z^{5/3} \exp \left (-\frac{\sigma_\text{beam}(z)^2}{z^2} k^2 - \frac{2z}{z_0} \right )
\end{multline}
which is now isotropic. Plugging this into Equation~\ref{eqn:sum-over-general-power-spectrum} and integrating over $t_i$ and $t_j$ yields\footnote{Here we use the fact that
\begin{equation}
    \int_{-\theta_\text{fov}/(2s)}^{+\theta_\text{fov}/(2s)} \int_{-\theta_\text{fov}/(2s)}^{+\theta_\text{fov}/(2s)}  e^{-v^2k^2(t_j - t_i)^2/2} dt_i dt_j  \approx  \sqrt{2\pi}\frac{\theta_\text{fov}}{ksv} 
\end{equation}
when $k \gg \frac{s}{\theta_\text{fov}v}$.
}
\begin{equation}
     P_\text{map}(k) = 
     \left [ \frac{f_\text{array}^2\theta_\text{fov} N_\text{scans}}{N_\text{per pixel}^2D_\text{map}} \right ] \left [ 8 \sqrt{2} \pi^2\frac{\Gamma(\tfrac{11}{6})}{\Gamma(\tfrac{1}{3})} \frac{g_\nu^2 \sigma_{\rho, \text{base}}^2 \theta_\text{fov}}{svr_0^{2/3}} k^{-14/3} \int dz \, z^{5/3} \exp \left (-\frac{\sigma_\text{beam}(z)^2}{z^2} k^2 - \frac{2z}{z_0} \right ) \right ].
\end{equation}
For small, isotropic patches on the sky, the angular power spectrum $P(k)$ and the multipole power spectrum $C_\ell$ are equivalent such that $C_\ell = P(\ell)$ (see Appendix~\ref{appendix:k-to-ell}). Thus, the above expression gives us our final map multipole power spectrum
\begin{equation}
\label{eqn:final-spectrum}
     C_\ell(k) = 
     128 \sqrt{2} \pi^2 \frac{\Gamma(\tfrac{11}{6})}{\Gamma(\tfrac{1}{3})} \frac{f_\text{sky}}{\theta_\text{fov}^2T_\text{survey} } \frac{g_\nu^2 \sigma_{\rho, \text{base}}^2}{vr_0^{2/3}} k^{-14/3} \int dz \, z^{5/3} \exp \left (-\frac{\sigma_\text{beam}(z)^2}{z^2} k^2 - \frac{2z}{z_0} \right )
\end{equation}
in terms of the sky fraction $f_\text{sky} = D_\text{map}^2 / 16$.

\section{Relating spherical harmonics and flat angular spectra}
\label{appendix:k-to-ell}

The angular covariance of an isotropic random spherical function

\begin{equation}
    f(\theta, \phi) = \sum_{\ell=0}^\infty \sum_{m=-\ell}^{\ell} a_{\ell m} Y_{\ell m} (\theta, \phi)
\end{equation}
is given by
\begin{equation}
\label{eqn:cmb-var}
    C(\theta) = \frac{1}{4 \pi} \sum_0^\ell (2\ell + 1) C_\ell P_\ell(\cos \theta)
\end{equation}
where $C_\ell = \langle a^*_{\ell m} a_{\ell m} \rangle$.
For small $\theta$ and large $\ell$ we have\footnote{See Theorem 8.21.2 in \cite{szego1975orthogonal}.}
\begin{equation}
    P_\ell(\cos \theta) = \sqrt{\frac{\theta}{\sin \theta}} \left ( J_0 \left [ (\ell + \tfrac{1}{2}) \theta \right ] + \frac{\tfrac{1}{\theta} - \cot \theta}{8(\ell + \tfrac{1}{2})} J_1 \left [ (\ell + \tfrac{1}{2}) \theta \right ] \right ) + \mathcal{O}(\ell^{-2})
    \approx J_0 \left [ (\ell + \tfrac{1}{2} ) \theta \right ],
\end{equation}
where $J_0$ is the Bessel function of the first kind of order $0$.
Writing Equation~\ref{eqn:cmb-var} as a continuous integral we get 
\begin{equation}
\label{eqn:hankel}
    C(\theta) = \frac{1}{4 \pi} \int (2\ell + 1) C_\ell J_0 \left [ (\ell + \tfrac{1}{2}  ) \theta \right ] d\ell \approx \frac{1}{2 \pi} \int C_\ell J_0 (\ell \theta) \ell d\ell.
\end{equation}
The integral on the right hand side is a Hankel transform of order 0, which is related to the Fourier transform $\F{\cdot}$ as 
\begin{equation}
    2 \pi \int f(r) J_0 (kr) r dr = \F{f(r)}.
\end{equation}
We can then write
\begin{equation}
    C(\theta) = \frac{1}{(2 \pi)^2} \F{C_\ell}
\end{equation}
and hence
\begin{equation}
    \F{C(\theta)} = \frac{1}{(2 \pi)^2} \F{\F{C_\ell}} = C_\ell.
\end{equation}
The Wiener-Khinchin theorem tells us that the autocorrelation of a process is the Fourier transform of the angular power spectrum of that process, and so for small $\theta$ and large $\ell$ the multipole power spectrum $C_\ell$ and locally flattened two-dimensional angular power spectrum $P(\ell) = \F{C(\theta)}$ are equivalent.

\end{document}